\documentclass{NAV}

\usepackage{multicol}
\usepackage{subcaption}

\begin{document}

\title[Towards Simple Machine Learning Baselines for GNSS RFI Detection]{Towards Simple Machine Learning Baselines for GNSS RFI Detection}

\author[V. Ivanov, R. C. Wilson, M. Scaramuzza]{Viktor Ivanov$^{1}$, Richard. C. Wilson$^{1}$, Maurizio Scaramuzza$^{2}$}

\address{\add{1}{Department of Computer Science, University of York}
\add{2}{Skyguide Swiss Air Navigation Services Ltd}}

\begin{abstract}
Machine learning research in GNSS radio frequency interference (RFI) detection often lacks a clear empirical justification for the choice of deep learning architectures over simpler machine learning approaches. In this work, we argue for a change in research direction-from developing ever more complex deep learning models to carefully assessing their real-world effectiveness in comparison to interpretable and lightweight machine learning baselines. Our findings reveal that state-of-the-art deep learning models frequently fail to outperform simple, well-engineered machine learning methods in the context of GNSS RFI detection. Leveraging a unique large-scale dataset collected by the Swiss Air Force and Swiss Air-Rescue (Rega), and preprocessed by Swiss Air Navigation Services Ltd. (Skyguide), we demonstrate that a simple baseline model achieves 91\% accuracy in detecting GNSS RFI, outperforming more complex deep learning counterparts. These results highlight the effectiveness of pragmatic solutions and offer valuable insights to guide future research in this critical application domain.
\end{abstract}


\maketitle

\section{INTRODUCTION}
Radio frequency interference (RFI) poses significant challenges to the use of Global Navigation Satellite Systems (GNSS) for navigation and surveillance in aviation \citep{eurocontrol2021does, scaramuzza2016empirical, truffer2017jamming, liu2020characterization, jonavs2019detection, ala2019detecting, mehr2022detection}. Aircraft operating under Instrument Flight Rules (IFR) increasingly rely on GNSS as a primary means of navigation during all phases of flight such as departure, en route, arrival, and landing. However, near the ground, GNSS signals are highly susceptible to low power jamming, both intentional and unintentional, due to low strength of the desired signal at the receivers antenna. Even low-power or distant jammers can significantly degrade GNSS performance, threatening the reliability of safety-critical aviation applications. These interferences can lead to unacceptable performance degradation, necessitating the need for advanced methods and systems that enable continuous and automatic detection of GNSS RFI.

At the 14th International Civil Aviation Organization (ICAO) Air Navigation Conference (AN-Conf/14) in 2024, concerns about GNSS jamming (and spoofing) have been increasingly acknowledged \citep{wp063e, wp151en, wp207en}. States are urged to establish monitoring and reporting mechanisms to mitigate these growing risks. The rise in GNSS RFI incidents, particularly in and around conflict zones, has further highlighted the urgency of this issue. Such interference often extends far beyond the targeted regions, adversely affecting GNSS receiver performance in unintended areas.

The integration of large-scale data processing presents a unique opportunity to address these challenges. Air navigation service providers (ANSPs) and airspace operators generate large amounts of high-quality communication, navigation, surveillance (CNS), and air traffic management (ATM) data. Leveraging these data streams through machine learning systems can enable automated decision-making by extracting valuable insights and offering statistically informed solutions to problems that traditional engineering techniques struggle to resolve.

Real-world aviation operations continuously produce extensive GNSS-related measurements. Efficiently detecting anomalies in these time-series datasets is critical for monitoring GNSS conditions in near real-time. This capability is essential for mitigating potential risks associated with GNSS RFI promptly. As GNSS-based navigation becomes increasingly integral to aviation operations, continuous signal interference detection is critical to maintaining reliable navigation and sustaining airspace capacity in modern air traffic environments.

A high-performing GNSS RFI detection model should be able to generalize well to unknown anomalies while learning complex nonlinear temporal patterns of the aircraft measurements' expected characteristics. GNSS timeseries have complex nonlinear temporal dependencies with other aircraft onboard signals. Additionally, GNSS RFI anomalies are highly infrequent and manually identifying and annotating these patterns is extremely labor-intensive. However, in reality, we might have a limited set of observations (e.g. flight recordings) that have been annotated as normal or abnormal, in addition to a huge set of unlabeled examples. Thus, we might be able to take effective advantage of such labeled data to detect anomalies through supervised machine learning methods.

Detecting GNSS RFI has increasingly attracted the attention of researchers applying deep learning methods, inspired by their success in domains like computer vision. While these models are capable of capturing complex temporal dependencies, their application is often motivated more by architectural and overall solution novelty than by demonstrable performance gains over classical, interpretable machine learning approaches. This reflects a broader concern in time series anomaly detection, where state-of-the-art deep models are frequently introduced without rigorous empirical validation, and often fail to significantly outperform robust, well-understood baseline methods \cite{sarfraz2024position}.

This work emphasizes the practical value of simplicity in GNSS RFI detection by demonstrating that well-designed, interpretable machine learning baselines can rival or exceed the performance of more complex deep learning models. Leveraging large-scale flight data collected from GPS and Attitude Heading Reference Systems (AHRS), we develop and evaluate efficient models capable of detecting interference events in near real-time. Our approach involves training on historical flight data containing both normal and anomalous conditions, and applying the resulting models to new, unseen recordings. The results challenge the prevailing assumption that increasing model complexity is necessary for effective RFI detection, showing instead that thoughtfully engineered baselines can offer both robustness and transparency in operational environments — a view consistent with recent critical evaluations in time series anomaly detection literature \cite{sarfraz2024position}.

We developed a scalable framework to detect GNSS RFI in Switzerland using a unique dataset from the Helicopter Recording Random Flights (HRRF) project, conducted by Skyguide, the Swiss Air Force, and Rega \citep{scaramuzza2016empirical, scaramuzza2015gnss, scaramuzza2014rfi, scaramuzza2019investigation, truffer2017jamming}. Data was collected over several years from about thirty helicopters operated by the Swiss Air Force and Rega, covering large portions of Switzerland under normal operating conditions. Low flight altitudes typical of these operations increase their susceptibility to low power GNSS RFI compared to higher altitude flights. GPS carrier-to-noise ratio measurements, combined with information from the Flight Management System (FMS) and Attitude and Heading Reference System (AHRS), enable effective statistical learning to identify RFI exposures.

While lab studies are critical for understanding GPS jamming, they often fail to replicate real world conditions \citep{truffer2017jamming}. Our dataset is unique, containing field trial recordings from military and civilian aircraft engaged in live jamming exercises conducted in Switzerland. These recordings demonstrate the impact of jamming on various GPS receivers and serve as real-world anomalies to test our machine learning model. Experimentally, our method detects GNSS RFI with 91\% accuracy. Extensive empirical studies demonstrate that the proposed method outperforms state of the art deep learning methods.

Similar to \citep{ivanov2024deep}, our method also addresses a key limitation of most of the existing approaches for GNSS RFI detection, especially the ones based on ADS-B consisting of that for achieving high detection accuracy, interference has to be large enough to totally disrupt the reception of GPS signals or remarkably deteriorate the position accuracy. Our approach enables detection of GNSS RFI in the absence of total GPS signal loss i.e. while the receiver is still able to determine a position which means RFI sources with low power or at larger distance could be detected.

\textbf{The main contributions of our work are:}
\begin{itemize}
\item Simple and effective machine learning baseline for near real-time detection of GNSS RFI that performs on par or better than state of the art deep learning methods, thus challenging the efficiency and effectiveness of increasing model complexity to solve GNSS RFI detection problems.
\item An evaluation on a real-world large scale aircraft data from Swiss Air Force and Rega, preprocessed by Skyguide, collected onboard and containing flight recordings impacted by a real jammer. The experimental results indicate that our system successfully detects potential GNSS RFI with 91\% accuracy. Extensive empirical studies demonstrate that the proposed method outperforms state of the art deep learning methods.
\end{itemize}

\section{RELATED WORK}
Before introducing our method, we briefly review previous approaches to GNSS RFI detection using machine learning.

A method outlined in \citep{liu2021gnss} provides a machine learning solution for detecting GNSS RFI that is based on ADS-B data. The authors used neural networks that learn from ADS-B data and generate a classification outcome indicating if the aircraft has been jammed. Navigation Integrity Category (NIC) is one of the primary raw features used by these models and RFI events from Cypriot airspace are used for positive examples. Due to the limited features in ADS-B messages, a key limitation of most of these approaches is that often interference has to be large enough to totally disrupt the reception of GPS signals or remarkably deteriorate the position accuracy which is not useful for situations where the interference impact is weak. 

Another line of research \citep{guo2021gnss, morales2019jammer, swinney2021gnss, mehr2022detection, ebrahimi2023deep, mehr2024deep} focuses on approaches to interference detection based on convolutional neural networks (CNNs) that learn from visual time-frequency representation of the received GNSS signal. Many of these approaches are based on artificially generated datasets that are not well representative for real-world flight measurements' dynamics. Last but not least, recorded aircraft measurements impacted by real jamming are hardly available which makes it difficult for applying conventional supervised machine learning methods and performing robust performance evaluation. 

The work of \citep{ivanov2024deep} presents a deep temporal semi-supervised one-class classification method for GNSS RFI detection developed and tested based on a large-scale real-world aircraft measurements data from  from Swiss Air Force and Swiss Air-Rescue (Rega), preprocessed by Swiss Air Navigation Services Ltd. (Skyguide), containing flight recordings impacted by real jamming which enables a rigorous assessment of the whole solution. 

\citep{mehr2024dual} introduces a dual-stage network architecture approach that operates at raw signal level (In-phase/Quadrature signal samples) and consists of a Long Short-Term Memory based autoencoder for GNSS interference detection and Convolutional Neural Networks for classification.

Overall, such approaches aim to leverage deep learning models for automatically learning a latent representations of GNSS timeseries data. Those representations are supposed to enable high detection performance. We argue that simple representations not requiring complex deep learning methods can perform equally or better.

\section{Simple Machine Learning Baseline for GNSS RFI Detection}
Recent approaches to GNSS RFI detection often emphasize algorithmic innovation, frequently adopting complex deep learning architectures. However, such complexity is not always matched by demonstrable gains in performance. Inspired by critical perspectives in time series anomaly detection research (e.g., \cite{sarfraz2024position}), we explore an alternative path: emphasizing simplicity, domain-aligned feature design, and robust evaluation. We present a lightweight machine learning baseline that surpasses a number of contemporary deep learning methods in identifying GNSS RFI. Importantly, this baseline also provides insights into the structure of the detection task itself. Our contribution lies in rigorously applying established techniques, crafting interpretable features tailored to GNSS behavior, and highlighting that well-grounded models can yield both performance and understanding.

\subsection{The GNSS RFI Detection Problem}
Carrier to noise ratios of all tracked satellites are the primary factors used for GNSS RFI detection in our work. Those are augmented with additional measurements such as heading, roll and pitch as well as ground speed and true air speed measurement. 

\begin{figure}[h]
\begin{center}
\includegraphics[width=8cm]{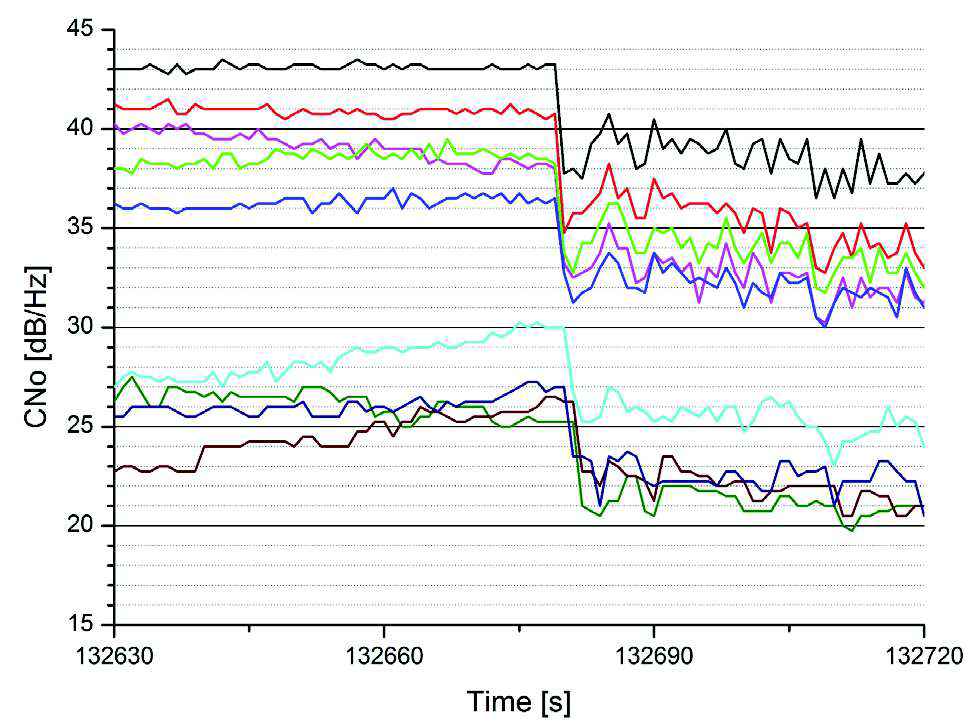}
\end{center}
\caption{RFI on a Stationary GNSS Receiver.}
\label{fig:1}
\end{figure}

\subsubsection{Interference on a Stationary GPS Receiver} Analyzing statistical properties of C/No values might reveal a potential radio frequency interference \citep{scaramuzza2014rfi}. Since RFI affects the entire GPS receiving antenna equally, all C/No ratios should theoretically decrease by a near constant level when exposed to interference \citep{scaramuzza2014rfi}. This hypothesis is empirically validated by \citep{scaramuzza2014rfi} where in a laboratory environment a real interference is assessed on a stationary GPS receiver which is illustrated in Figure \ref{fig:1}. When interference started all C/No measurements dropped by five dB.

\subsubsection{Interference Detection on a Live GNSS Receiver} 
\begin{figure*}
\begin{multicols}{2}
    \includegraphics[width=6cm]{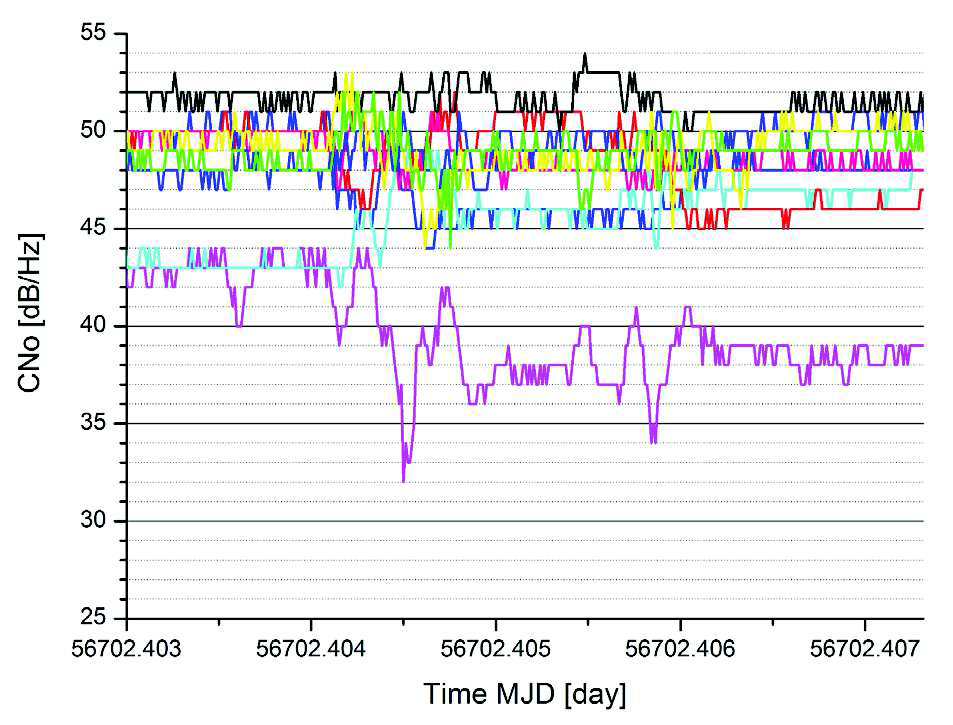}\caption{Helicopter Maneuver Affecting GNSS Satellite's C/No.}\label{fig:2}\par
    \includegraphics[width=6cm]{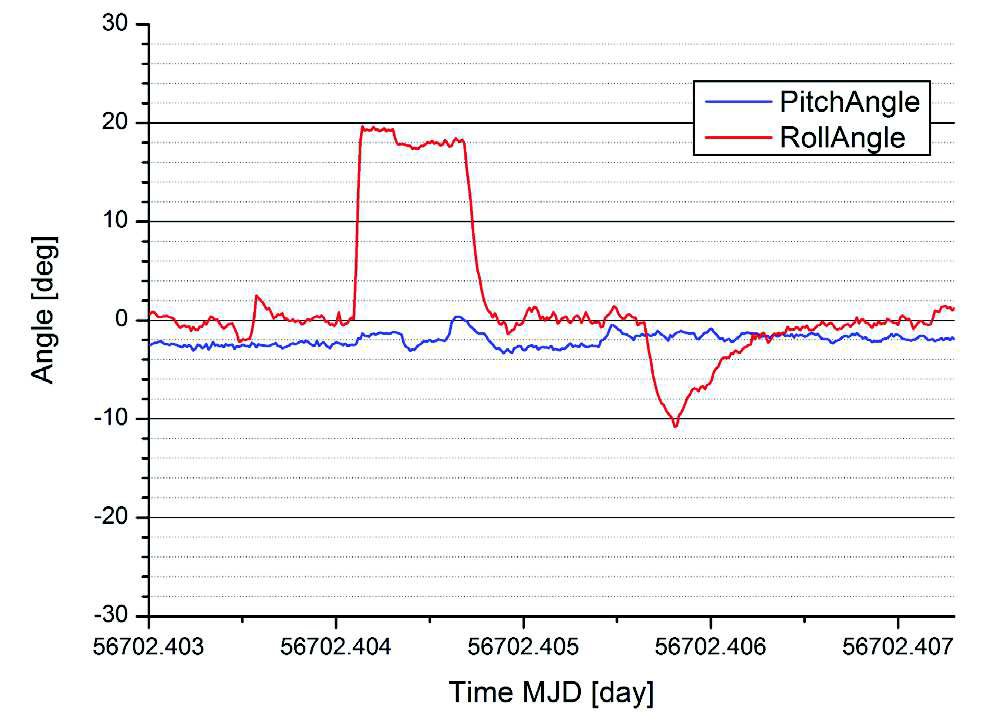}\caption{Roll and Pitch Angles Affecting the Carrier to Noise Ratio Shown in Figure \ref{fig:2}.}\label{fig:3}\par
\end{multicols}
\end{figure*}

We could address in an analogous way the detection of interference for GNSS receiver running in a live setting, however, there are some challenges to be aware of. In contrast to the laboratory example in figure \ref{fig:1}, a flying aircraft that approaches an interference source would typically be impacted gradually and the carrier to noise ratio would smoothly decrease \citep{scaramuzza2014rfi}. Second, carrier to noise is impacted by the satellite positions relative to the antenna \citep{scaramuzza2014rfi}. Figure \ref{fig:2} illustrates a real scenario assessed by \citep{scaramuzza2014rfi} where changes in the aircraft's attitude are shown to have an impact on some C/No values. Considering that some C/No ratios are decreasing while others are increasing, the drop at 56702.404 time cannot be attributed to radio frequency interference \citep{scaramuzza2014rfi}. According to the measured roll and pitch angles, a maneuver has been performed influencing the C/No ratios by different values (Figure \ref{fig:3}).

The factors such as multipath, antenna gain and atmosphere, that can affect the carrier to noise values in addition to an RFI and how these are treated in the data treatment are described in \citep{scaramuzza2014rfi}.

\subsection{Statistical Learning Problem Definition}
Let \( X \) be a set of \( N \) time series representing recorded flights and $X_i = \{ x_{i,1}, x_{i,2}, \dots, x_{i,T} \}$ represents a particular flight recording $\text{for } i = 1, \dots, N,$ where \( T \) is the length of the flight \( X_i \). Each element \( x_{i,t} \) $\in \mathbb{R}^M$ in a time series is a vector encoding aircraft measurements at a particular time epoch during a flight i.e., carrier to noise (C/No$_{1}$..C/No$_{32}$), heading, roll, pitch, ground speed (velocity over ground measurement where no wind effects taken into account), true air speed measurement (velocity relative to air). For each time series \( X_i \) we have a discrete label $y_i \in \mathcal{Y}$ where $\mathcal{Y} = \{0, 1\}$ and $0$ represents normal flight measurements and $1$ represents abnormal i.e., potential RFI. The label is assigned for the whole timeseries i.e., for the whole flight recording. The goal is to learn a decision function \( f : \mathbb{R}^T \to \mathcal{Y} \) from a labeled training set of flight recordings \( \{ (X_i, y_i) \}_{i=1}^N \) that minimizes the classification error on unseen flight recordings. Formally, this objective can be expressed as finding:
\begin{equation}
     f^* = \arg \min_f \frac{1}{N} \sum_{i=1}^N f(X_i) \neq y_i
    \label{eq:6}
\end{equation}

\subsection{Proposed simple and effective baseline}
The core idea behind our approach is to assess the C/No values distribution of each tracked GPS satellite \citep{scaramuzza2015gnss,scaramuzza2014rfi} together with heading, roll and pitch measurements as well as ground speed, i.e. velocity over ground measurement (no wind effects taken into account) and true air speed measurement i.e., velocity relative to air, taking into account that the GPS receiving antenna is entirely impacted by the same level of interference, an RFI occurrence might cause the C/No to decrease by a near constant amount at each individual time step \citep{scaramuzza2015gnss,scaramuzza2014rfi}. It is possible to determine whether a systematic decrease pattern of all C/No is present, indicating a potential RFI, or not. In our context man-made RFI is detected. It is understood, that RFI is not the only electromagnetic type of interference to have an adverse impact on the quality of the GNSS signals. Only non-intentional and intentional man-made RFI is subject of our analysis i.e. multipath and natural RFI are out of scope.

\begin{figure}[h]
\begin{center}
\includegraphics[width=8cm]{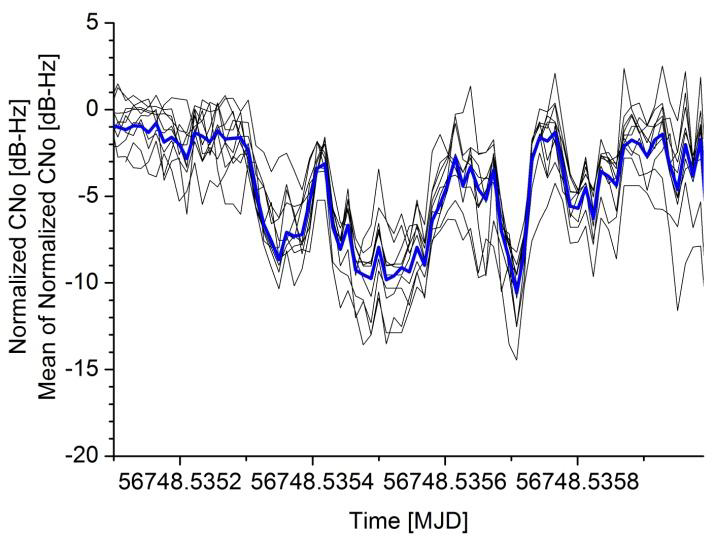}
\end{center}
\caption{Situation where a potential RFI might be present. The mean value of the normalized C/No decreases by up to 10 dB-Hz. The standard deviation remains similar compared to situations without RFI.}
\label{fig:10}
\end{figure}

To illustrate the above reasoning, a typical event is shown in Figure \ref{fig:10}. The thin black lines represent the normalized C/No of all other tracked satellites. A normalized C/No is derived by taking into account the influence of all the parameters mentioned above and calculating them into a single value per point in time \citep{scaramuzza2015gnss,scaramuzza2014rfi}. The blue line represents the mean value of normalized C/No. The time span is 90 seconds. During this period, the mean value falls by up to 10 dB while the standard deviation of the normalized C/No remains similar to the situation when no RFI is present. This is a strong indication that all GPS signals are affected by the same amount of perturbation. So this effect could be explained by a potential RFI \citep{scaramuzza2015gnss}.

Following that reasoning, we derive a set of logical feature functions based on sensor range deviation dynamics for detecting potential GNSS RFI. The range of sensor values observed during normal operation can be useful in identifying out-of-distribution aircraft measurement signals. Anomalies in GNSS time series data can occur when the sensor values deviate from their usual range. Therefore, if the sensor values in a test GNSS timeseries point fall outside the observed range, it may indicate the presence of an anomaly, and thus a potential RFI. Formally, this is defined as:

\begin{equation}
    f({\hat{x}}_t) = \begin{cases} 
                        0 & if \hspace{5pt} \hat{x}_t \in [\min(X), \max(X)] \\
                        1 & otherwise
                     \end{cases}.
    \label{eq:5}
\end{equation}

This represents a minimum level of detection performance that any advanced method should be able to surpass. Let $\textbf{x} = \{x_1, x_2, ..., x_n\}$ denotes a timeseries of length $n$ representing a given sensor values during a flight recording. We derive a set of ten core feature functions capturing basic sensor range deviation dynamics based on first and second order signal statistics:

\begin{itemize}

\item \textbf{Sum Values:} Calculates the total sum of all values in the time series. This feature provides insight into the overall magnitude of the series and can highlight trends or biases toward higher or lower values.

\[
\text{Sum} = \sum_{i=1}^{n} x_i
\]

\item \textbf{Median:} Computes the median of the time series, which is the middle value when all values are sorted. This is a measure of central tendency and is robust to outliers compared to the mean.

\[
\text{Median} =
\begin{cases}
x_{\frac{n+1}{2}}, & \text{if } n \text{ is odd}, \\
\frac{x_{\frac{n}{2}} + x_{\frac{n}{2} + 1}}{2}, & \text{if } n \text{ is even}.
\end{cases}
\]

\item \textbf{Mean:} Computes the average of all values in the time series. It represents the central tendency of the data but is sensitive to extreme values (outliers).

\[
\text{Mean} = \frac{1}{n} \sum_{i=1}^{n} x_i
\]

\item \textbf{Length:} Measures the total number of observations or data points in the time series. This is useful for understanding the temporal resolution and overall span of the data.

\[
\text{Length} = n
\]

\item \textbf{Standard Deviation:} Calculates the standard deviation of the values in the time series, representing the degree of variation or spread around the mean. A higher standard deviation indicates greater variability in the data.

\[
\text{Standard Deviation} = \sqrt{\frac{1}{n} \sum_{i=1}^{n} \left(x_i - \text{Mean}\right)^2}
\]

\item \textbf{Variance:} Computes the variance of the time series, which is the square of the standard deviation. It quantifies the dispersion of the values from their mean and can help identify consistency or irregularities in the data.

\[
\text{Variance} = \frac{1}{n} \sum_{i=1}^{n} \left(x_i - \text{Mean}\right)^2
\]

\item \textbf{Root Mean Square:} Calculates the root mean square of the time series values. This feature combines both the magnitude and variability of the values, emphasizing larger deviations.

\[
\text{Root Mean Square} = \sqrt{\frac{1}{n} \sum_{i=1}^{n} x_i^2}
\]

\item \textbf{Maximum:} Determines the highest value in the time series. This feature highlights peaks in the data and can indicate upper bounds or extreme events.

\[
\text{Maximum} = \max_{i} x_i
\]

\item \textbf{Absolute Maximum:} Identifies the maximum absolute value in the time series, regardless of sign. It highlights the largest deviation from zero, capturing both large positive and negative extremes.

\[
\text{Absolute Maximum} = \max_{i} |x_i|
\]

\item \textbf{Minimum:} Determines the lowest value in the time series. This feature highlights troughs in the data and can indicate the lowest bounds or significant drops.

\[
\text{Minimum} = \min_{i} x_i
\]

\end{itemize}

The feature functions are computed for each raw input signal available in the flight recording timeseries.

Maximum, absolute maximum, and minimum feature functions are generally considered non-robust, as they can be disproportionately affected by noise. A common approach to mitigate this issue is to use the 95th and 5th percentiles, as well as the robust range (i.e., 95th percentile minus 5th percentile), which are less sensitive to outliers. However, in our case, we observed superior model performance when using the maximum, absolute maximum, and minimum features instead of percentile-based alternatives.

\textbf{Simple Machine Learning Baseline} Given the outlined feature functions forming a feature transformation \( \phi : \mathbb{R}^M \to \mathbb{R}^F \) where \( F \) ($M \times 10$) is the dimension of the new feature space, each flight recording \( X_i \) is transformed into a feature vector \( \phi(X_i) \in \mathbb{R}^F \). Then, any supervised machine learning model can be utilized to optimize the objective function in Equation \ref{eq:6}. In our work we employ two of the most widely used machine learning model baselines: logistic regression \citep{cox1958regression} as a representative linear model and Light GBM \citep{ke2017lightgbm} as a representative non-linear model. Data are scaled by removing the mean and scaling to unit variance.

\section{EXPERIMENTAL EVALUATION}
In this section, we describe the evaluation of the proposed machine learning approach. Our main objective is to study how effective our approach is for GNSS RFI detection based on a real-world dataset containing onboard measurements of jammed aircrafts.

\subsection{Data} 
\begin{figure*}
\begin{multicols}{2}
    \includegraphics[width=6cm]{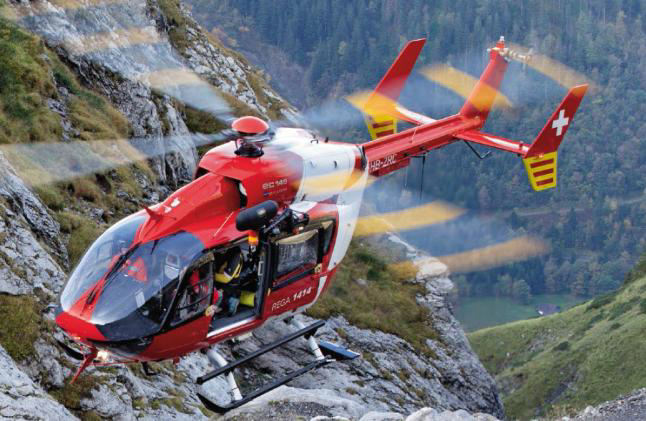}\caption{EC145 of the HEMS operator REGA. The GPS antenna is installed on the top of the fin in front of the strobe light (Courtesy REGA).}\label{fig:5}\par
    \includegraphics[width=6cm]{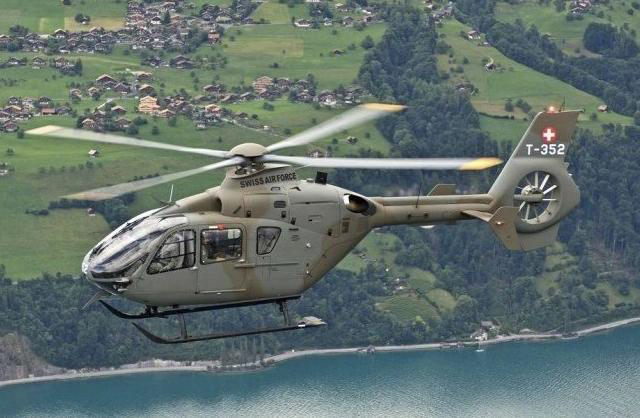}\caption{EC635 of the Swiss Air Force. The GPS antenna is mounted analogous to the EC145 (Courtesy VBS).}\label{fig:6}\par
\end{multicols}
\end{figure*}

This research utilizes data provided by Skyguide, consisting of over 54,000 recorded flights gathered during the Helicopter Recording Random Flights (HRRF) project. Data recorders were installed on approximately three dozen helicopters, including AW109SP, EC-145 models (Figure \ref{fig:5}), and EC-635 (Figure \ref{fig:6}), operated by Rega and the Swiss Air Force \citep{scaramuzza2016empirical, scaramuzza2015gnss, scaramuzza2014rfi, scaramuzza2019investigation, truffer2017jamming, scaramuzza2017quality}. Over six years, data from the Global Positioning System (GPS), Attitude and Heading Reference System (AHRS), and Flight Management System (FMS) was logged under normal operating conditions, covering large portions of Switzerland. Due to their low-altitude flight profiles, these helicopters are more likely to encounter GNSS RFI compared to higher-altitude aircraft. Such exposures can be detected through recorded C/No values or position losses \citep{scaramuzza2015gnss,scaramuzza2014rfi}. The dataset's comprehensive nature further supports effective statistical learning by incorporating diverse input features.

\begin{figure}[h]
\begin{center}
\includegraphics[width=6cm]{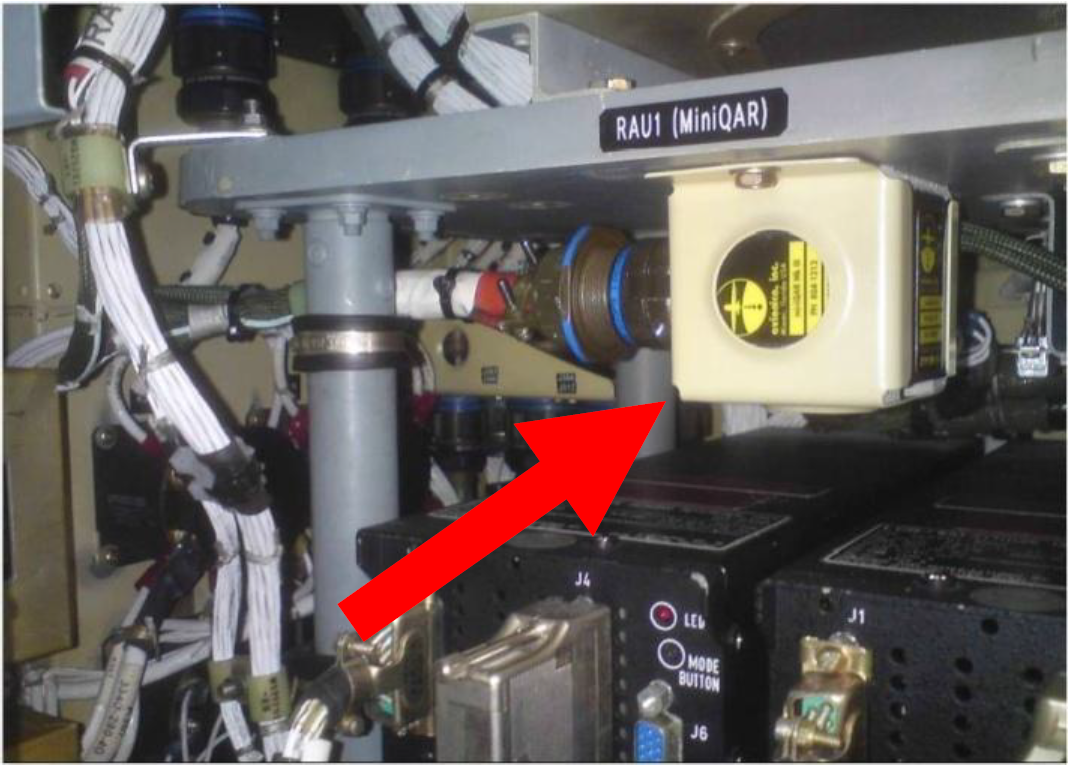}
\end{center}
\caption{Installed Avionica mQAR in Swiss Air Force and Rega Helicopters (red arrow).}
\label{fig:7}
\end{figure}

\subsubsection{Installation} The technical setup, as described in \citep{scaramuzza2015gnss,scaramuzza2014rfi}, involves connecting a mini Quick Access Recorder (mQAR) to the helicopter’s ARINC bus or RS-232 interface, depending on the aircraft’s architecture. The mQAR, a lightweight and compact unit (Figure \ref{fig:7}, red arrow), automatically records data from power-on to power-off of the main helicopter systems, requiring no pilot or ground crew interaction. Data is stored on a Secure Digital (SD) memory card, capable of holding several weeks of flight data under normal conditions. Ground staff at each helicopter base were instructed to download the data and upload it to a shared storage system every two to four weeks, ensuring consistent data availability for analysis.

\subsubsection{Recorded Data} The dataset comprises extensive flight data. For the EC-145 and EC-635 models, data is available from the GPS, AHRS, and FMS. The GPS data includes satellite position, GPS position, horizontal and vertical integrity limits (HIL and VIL), figures of merit (FOM), pseudo ranges (PR), pseudo range rate (PRR), C/No ratios, and various status indications. The AHRS data encompasses roll, heading, and pitch measurements. Both GPS and AHRS systems operate at a sampling rate of 1 Hz, ensuring detailed temporal resolution for analysis.

\subsubsection{Interference Detection Input} The primary measurement used for GNSS interference detection is the carrier-to-noise ratio (C/No) of each tracked satellite. Radio frequency interference affects all carrier-to-noise ratios. To enhance the detection process, the C/No data (C/No$_{1}$..C/No$_{32}$) is supplemented with additional measurements, including heading, roll, pitch, yaw, ground speed (velocity over ground, without accounting for wind effects), and true airspeed (velocity relative to air). This combination provides a much richer dataset for analysis compared to conventional ADS-B setups, offering greater context for interference detection.

\subsubsection{Interference Labels} Each flight recording in the dataset is classified as either normal or abnormal (potential RFI) by Skyguide, following the method described by \citep{scaramuzza2015gnss, scaramuzza2014rfi}. The dataset also includes two flight recordings from real-world field trials with live jamming scenarios involving military and civilian aircraft. These trials were conducted by Skyguide and the Swiss Air Force \citep{truffer2017jamming}. The recorded data provide valuable insights into how real-world jamming impacts different GPS receivers across various aircraft types. Additionally, the jamming trials serve as a basis for testing the developed detection models.

\subsubsection{Training Data} For model training we use the flight recordings of 17 of the helicopters available in the dataset provided by Skyguide with RFI labels provided by Skyguide, derived from the method developed by \citep{scaramuzza2015gnss, scaramuzza2014rfi}. We explicitly exclude all flights of the two helicopters participated in the jamming trials to prevent potential overfitting on specific aircraft. We have 34,482 training flight recordings (timeseries) in total where the average number of epochs per flight (timeseries length) is 1,700 and there are $15\%$ anomalous flight recordings in the training data.

\subsubsection{Test Data} For model evaluation we use flight recordings of 3 of the helicopters (not part of the training set) including the two flights from the jamming trials. The test dataset has 7,888 flight recordings (timeseries) out of which 964 (12.2\%) are labeled as potential RFI. RFI labels are provided by Skyguide.

\subsection{Benchmark Models}
The following set of machine learning algorithms are used for evaluation and performance comparison:

\textbf{Simple Linear Baseline} We implement a linear baseline model based on the proposed feature functions i.e., logistic regression machine learning model \citep{cox1958regression}. Logistic regression is a supervised machine learning algorithm for binary classification, predicting the probability that an input belongs to one of two classes. It models the relationship between input features and the target using a sigmoid function, transforming linear predictions into probabilities between 0 and 1. A threshold (e.g., 0.5) is applied to assign class labels. The model is trained using cross-entropy loss to minimize prediction errors. While easy to implement and interpret, logistic regression performs best with linearly separable data and may struggle with complex, non-linear relationships.

\textbf{Simple Non-Linear Baseline} We also implement a non-linear baseline based on the proposed feature functions. We utilize LightGBM (Light Gradient Boosting Machine) \citep{ke2017lightgbm} -- a powerful and efficient gradient boosting framework specifically designed for speed and scalability in machine learning tasks. It employs a histogram-based approach to split finding, which reduces memory usage and computational cost. LightGBM supports both classification and regression tasks and excels in handling large-scale datasets with high-dimensional features. It features innovations such as leaf-wise tree growth, which optimizes for lower loss at the cost of potentially deeper trees, and categorical feature handling without one-hot encoding. These design choices make LightGBM particularly suitable for tasks requiring high accuracy and fast training times, such as time-series forecasting, ranking, and real-time predictions. In evaluation, its ability to balance performance and efficiency has made it a benchmark tool in modern machine learning research and applications.

Next we describe the deep learning architectures we test against our simple model baselines. For all deep learning models, data are scaled by removing the mean and scaling to unit variance. Subsequently, data is padded to ensure that all time series samples have the same length.

\textbf{ResNet} The ResNet (Residual Network) architecture, introduced by \citep{he2016deep}, addresses the degradation problem in deep neural networks by employing residual learning. Instead of learning the desired mapping directly, ResNet reformulates the problem to learn residual functions, which represent the difference between the input and output of a layer. Its key innovation is the use of skip connections, which bypass one or more layers, allowing gradients to flow more efficiently during backpropagation. This design enables the construction of very deep networks, achieving state-of-the-art performance on tasks like image classification and object detection. ResNet's scalability and robustness have made it a foundational model in deep learning research and applications. It is adapted for time series classification by treating the time series data as input sequences instead of images. The ResNet architecture consists of stacked residual blocks, where each block has convolutional layers with filters tailored to extract features from the time series data. These layers capture temporal patterns, trends, and interactions between variables. The residual connections help mitigate the vanishing gradient problem and enable deeper architectures to be trained effectively. The output of the residual blocks is passed to a global average pooling layer, which reduces the feature dimensions, followed by a fully connected layer and a softmax activation to classify the time series into predefined categories.

\textbf{InceptionTime} InceptionTime, introduced by \citep{ismail2020inceptiontime}, is a deep learning architecture specifically designed for time series classification. Inspired by the Inception module in convolutional neural networks, it employs multiple parallel convolutional filters with varying kernel sizes, enabling the capture of features across different time scales. By stacking these modules, InceptionTime achieves a high capacity to model diverse temporal patterns while maintaining computational efficiency. The architecture demonstrated state-of-the-art performance across a wide range of benchmark datasets, surpassing traditional methods and other deep learning models. Its adaptability and effectiveness have established it as a key model for time series analysis.

\textbf{FCN} The Fully Convolutional Network (FCN), proposed by \citep{wang2017time}, is a deep learning model tailored for time series classification. It utilizes only convolutional layers, eliminating the need for fully connected layers, which reduces the number of parameters and enhances computational efficiency. FCN captures local temporal dependencies and hierarchical feature representations using convolutional filters, coupled with global average pooling to condense information before the final classification layer. This architecture demonstrated competitive performance on time series benchmarks, setting the stage for convolution-based methods in time series classification tasks.

\textbf{LSTM-FCN} The LSTM Fully Convolutional Network (LSTM-FCN), introduced by \citep{karim2017lstm}, combines the strengths of convolutional neural networks (CNNs) and long short-term memory (LSTM) networks for time series classification. The architecture integrates fully convolutional layers to extract local and hierarchical features, alongside an LSTM module to model temporal dependencies and long-term patterns in the data. Global average pooling is applied to reduce dimensionality and enhance generalization before the classification layer. LSTM-FCN achieved state-of-the-art performance on diverse time series datasets, showcasing its ability to effectively capture both spatial and temporal characteristics of time series data.

\textbf{MiniRocket}, introduced by \citep{tan2022multirocket}, is an efficient and highly accurate method for time series classification. It builds on the Rocket framework but significantly reduces computational complexity by using a fixed set of random convolutional kernels and streamlined feature extraction. MiniRocket generates a small number of summary statistics, such as proportion of positive values, from the convolutions, which are then used as inputs for a simple linear classifier. Despite its simplicity and speed, MiniRocket achieves state-of-the-art accuracy on diverse time series datasets, making it a practical and scalable solution for time series analysis.

\textbf{mWDN} (multi-Window Dilated Network) is a deep learning architecture designed for time series forecasting and classification tasks \citep{wang2018multilevel}. It leverages multiple receptive fields through dilated convolutions across various window sizes, enabling the model to efficiently capture both short-term and long-term dependencies in time series data. By using multiple windows, mWDN enhances the extraction of temporal patterns while maintaining computational efficiency. This architecture is particularly effective for handling complex time series datasets, offering improved accuracy and scalability compared to traditional single-window models.

\textbf{MLSTM-FCN} (Multi-Layer Long Short-Term Memory - Fully Convolutional Network) is a hybrid deep learning architecture designed for time series classification tasks \citep{karim2019multivariate}. It combines the strengths of LSTM layers, which excel at capturing long-term temporal dependencies, with Fully Convolutional Networks (FCN) that efficiently extract local patterns and features through convolutional layers. The model uses an attention mechanism to emphasize the most relevant time steps, enhancing interpretability and performance. This synergy allows MLSTM-FCN to handle complex temporal dynamics effectively, making it a powerful choice for multivariate and univariate time series classification problems.

\subsection{Performance Metrics} We use five standard metrics for evaluating our method against the deep learning models: ROC AUC, Precision, Recall, F1 Score and Accuracy:

\begin{itemize}
  \item \textbf{ROC AUC} measures the model's ability to distinguish between classes by calculating the area under the ROC curve, which plots the Recall against the False Positive Rate (FP / FP + TN). An AUC of 1.0 indicates perfect classification, while 0.5 represents random guessing.
  \item \textbf{Precision} measures the proportion of correctly predicted positive instances out of all instances predicted as positive, focusing on the accuracy of positive predictions. It is calculated as TP / (TP + FP).
  \item \textbf{Recall} measures the proportion of correctly predicted positive instances out of all actual positive instances, focusing on the model's ability to capture all relevant cases. It is calculated as TP / (TP + FN).
  \item \textbf{F1 Score} is a metric that balances precision (accuracy of positive predictions) and recall (ability to find all positives) into a single performance measure. It is calculated as 2 * (Precision * Recall) / (Precision + Recall).
  \item \textbf{Accuracy} measures the proportion of correctly classified instances out of the total instances. It is calculated as (TP + TN) / Total Instances. It provides a general sense of how often the model makes correct predictions. However, accuracy on its own is not a sufficient metric for measuring performance in imbalanced classification problems such as the GNSS RFI detection problem. Accuracy does not account for the distribution of classes, and a classifier can achieve a high accuracy by simply predicting the majority class all the time — without actually learning anything meaningful. For example, in our case 85\% of the data belongs to the negative class and only 15\% to the positive (potential RFI) class. A naive classifier that always predicts the negative class will achieve 95\% accuracy— even though it completely fails to recognize the positive class. Thus, the whole range of provided metrics must be considered.
\end{itemize}

TP refers to true positive which is the outcome where the model correctly predicts the anomaly class, TN refers to true negative which is the outcome where the model correctly predicts the normal class, FP false positive which is the outcome where the model incorrectly predicts the anomaly class and FN refers to false negative which is the outcome where the model incorrectly predicts the normal class.

\subsection{Hardware}
We ran our experiments on a standard desktop workstation with a 12th Gen Intel(R) Core(TM) i9-12950HX 2.30GHz processor, 128GB RAM, NVIDIA RTX A5500 16GB GPU, a solid state drive storage, and running 64-bit Windows 11 Pro.

\subsection{Evaluation Results}
Table \ref{tab:1} shows the results on all flights from the three helicopters used for testing. The main comparison metric is ROC AUC score which is commonly used as a standard benchmark metric in imbalanced binary classification. Precision, Recall, F1 Score and Accuracy are also reported.

\begin{table}[h]
\makebox[\textwidth]{
\begin{tabular}{||l c c c c c||} 
 \hline
 \textbf{Method} & \textbf{ROC AUC} & \textbf{Precision} & \textbf{Recall} & \textbf{F1 Score} & \textbf{Accuracy} \\ [0.5ex] 
 \hline\hline
 \textbf{Simple Linear Baseline} & 0.86 & 0.51 & 0.48 & 0.50 & 0.88 \\ 
 \hline
 \textbf{Simple Non-Linear Baseline} & \textbf{0.91} & \textbf{0.74} & \textbf{0.38} & \textbf{0.50} & \textbf{0.91} \\ 
 \hline
 ResNet & 0.81 & 0.67 & 0.23 & 0.34 & 0.89 \\ [1ex] 
 \hline
 InceptionTime & 0.80 & 0.69 & 0.20 & 0.31 & 0.89 \\ [1ex] 
 \hline
 FCN & 0.80 & 0.69 & 0.20 & 0.31 & 0.89 \\ [1ex] 
 \hline
 LSTM-FCN & 0.80 & 0.64 & 0.22 & 0.33 & 0.89 \\ [1ex] 
 \hline
 MiniRocket & 0.83 & 0.69 & 0.29 & 0.41 & 0.90 \\ [1ex] 
 \hline
 mWDN & 0.82 & 0.68 & 0.23 & 0.34 & 0.89 \\ [1ex] 
 \hline
 MLSTM-FCN & 0.82 & 0.67 & 0.23 & 0.34 & 0.89 \\ [1ex] 
 \hline
\end{tabular}
}
\caption{Model Evaluation on all flights from the three helicopters used for testing.}
\label{tab:1}
\end{table}

Based on our test set the estimated average inference time of our method is 0.1 seconds per flight where the average number of epochs in a flight is 1700 considering that epoch sampling interval is 1Hz.

What we observe is that even a simple linear model trained on the proposed baseline feature set demonstrates superior performance compared to advanced deep learning systems, achieving higher ROC AUC and F1 scores. When a more advanced non-linear classifier is applied to the same feature set, it exhibits consistent dominance across all performance metrics.

\subsection{Discussion}
Recent critiques in the time series anomaly detection literature have highlighted the limitations of deep learning approaches when benchmarked against simple yet effective baselines \cite{sarfraz2024position}. In alignment with these findings, our results show that a range of advanced deep learning models applied to GNSS RFI detection were consistently outperformed by a straightforward machine learning baseline. This underperformance may be attributed to factors such as limited training data and overfitting to normal flight behavior. While deep learning methods have demonstrated strong performance in many domains, their added complexity and reduced interpretability offer little advantage in this safety-critical setting. Perhaps more concerning is that the widespread use of these type of models—without proper comparison to simple baselines—can give a misleading impression of progress. To address this, we propose a transparent and easily reproducible benchmark that establishes a reliable foundation for future research in GNSS RFI detection. This method offers a practical means of distinguishing between trivial and genuinely difficult anomalies—an essential step toward advancing robust and meaningful solutions. Our conclusions echo broader insights in unsupervised anomaly detection, where simple models have repeatedly rivaled or exceeded the performance of state-of-the-art deep learning methods \cite{sarfraz2024position}.

\textbf{Application to Vehicles on Ground} In principle, our approach could also be used for vehicles on the ground, not only for helicopters, but a poorer performance is to be expected. The multipath effects will be stronger from surrounding environments and are likely to be problematic. In the case of helicopters, the multipath effect is largely reduced, as data was only used if the velocity was larger than 10m/s. It can therefore be expected that multipath effects only occur for a short time period and therefore hardly influence the result (exceptions can be when the helicopter flies over a smooth water surface, but this was rarely the case). Furthermore, the helicopters usually fly higher above the ground if they are moving at more than 10m/s and therefore potential reflectors are further away from the receiving antenna. Depending on the correlator spacing of the receiver used, the multipath error collapses from a certain distance and therefore also contributes to a better result. In the conservative case of BPSK1 and 1 chip correlator spacing, the error is practically zero with a multipath delay of just over 400m. For a vehicle on the ground, it can be assumed that these two influences are stronger than for a helicopter. Especially when, for example, a car is moving slowly in city traffic and is surrounded by reflectors such as building walls and the ground, this method is likely to lead to degraded results.

\textbf{Training with Synthetic Jamming Signals} Considering that real-world jamming data is hard to acquire, users of our model could, in principle, use a software-based approach to add synthetically generated jamming signals to real measurements of GNSS, and use the modified GNSS measurements to train the model. In such a case the jammer has to be modelled and this highly depends on the environment (terrain, obstacles etc.), the jammer antenna characteristics, etc. and depending on the applied model, the result might be better or not.

\section{CONCLUSIONS}
We presented a simple machine learning approach for detection of GNSS RFI from large amounts of aircraft data. Extensive empirical studies demonstrate that the proposed method outperforms state of the art deep learning models. We hope our work will help improve the research efforts on GNSS RFI detection by encouraging caution in the premature adoption of complex tools for the task.

\section*{Acknowledgements}

This research was made possible by the Swiss Air Navigation Service Provider - Skyguide, Swiss Airforce and Swiss Air-Rescue (Rega).

\end{document}